\documentclass[11pt,twoside,a4paper]{article}
\usepackage[affil-it]{authblk}
\usepackage{graphicx}
\usepackage{amsmath,amsbsy}
\newcommand{\bs}[1]{\ensuremath{\boldsymbol {#1}}}
\usepackage{dcolumn} %align on decimal point
\usepackage[novbox]{pdfsync}   %ohne novbox macht es tabularx kaputt
\usepackage{siunitx}
\usepackage{hyperref}
\usepackage[square,numbers,compress]{natbib}
\usepackage{subfigure}

\begin{document}

\title{Simulating the dynamics of complex plasmas}
\author{M. Schwabe}
\affil{Department of Chemical and Biomolecular Engineering,
  University of California, Berkeley, CA 94720, USA\\Max Planck Institute for extraterrestrial Physics, PO Box 1312, Giessenbachstr., 
   85741 Garching, Germany}

\author{D. B. Graves}
\affil{Department of Chemical and Biomolecular Engineering,
  University of California, Berkeley, CA 94720, USA}

\date{}

\maketitle

\begin{abstract}
Complex plasmas are low temperature plasmas that contain
micro\-meter-sized particles in addition to the neutral gas particles
and the ions and electrons that make up the plasma. The microparticles
interact strongly and display a wealth of collective effects.
Here, we report on linked numerical simulations that reproduce many of
the experimental results of complex plasmas. We model a capacitively coupled plasma with a fluid code written for the commercial package COMSOL. The output of this model is used to
calculate forces on microparticles. The microparticles are modeled
using the molecular dynamics package LAMMPS, which we extended to
include the forces from the plasma. Using this method, we are able to reproduce void formation, the
separation of particles of different sizes into layers, lane
formation, vortex formation and other effects.
\end{abstract}

\section{Introduction}

In 1994, three research groups succeeded in forming crystals composed of small
particles in a low temperature plasma, called ``plasma crystals''
\cite{Thomas1994,Chu1994,Hayashi1994}. These systems
ignited the interest of scientists world-wide, and the research on
them has since increased dramatically. In general, charged
microparticles embedded in a weakly ionized 
plasma are called ``complex plasmas'' in analogy to complex liquids --
soft matter systems in the liquid form \cite{Morfill2009}. The term
``complex plasma'' sets these systems apart from the naturally
occurring dusty plasmas, even though this distinction is not always
made. 

Complex plasmas are of great interest as the microparticles can become the dominant species
regarding energy and momentum transport \cite{Morfill2009}, and they
display a multitude of collective effects that occur in other systems
as well. Complex plasmas can often serve as model for these other systems. 

Numerical simulations of complex plasmas are useful to check the
theoretical understanding, to make estimates of plasma parameters that
are not easily accessible in experiments, and to predict parameters
for future investigations. Three main approaches to simulations of complex plasmas exist, along
with variations and hybrid models of these approches. 

Firstly, the plasma is modeled as a fluid. If charging processes are
investigated, applying a fixed potential at the surface
of a stationary particle suffices \cite{Melandso1995}. In order to simulate
larger dynamic systems, the microparticles are treated as an additional fluid in a
simulation of the whole complex plasma
\cite{Akdim2001,Akdim2003a,Akdim2003b,Gozadinos2003,Land2010}. \citet{Akdim2003}
introduced tracer particles into such a model and were able to
reproduce the vortices that often form in weightless complex plasmas.

These fluid models suffer from the problem that the time scale of the microparticle dynamics is much 
larger than that of the plasma, and the simulation typically switches
periodically between advancing the plasma fluid and advancing the
microparticle fluid, until an equilibrium is reached
\cite{Land2006,Land2010}. The microparticles and the plasma are coupled via the
Poisson equation and the forces acting on the microparticles. 

The second approach to numerical simulations of complex plasmas is to
use Particle-In-Cell (PIC) methods, often coupled with Monte Carlo
(MC) simulations, to study the movement of ions and
electrons in the vicinity of microparticles \cite{Block2012}. This
approach is very effective in investigating ion drag effects, wakes and the charging
process of microparticles
\cite{Choi1994a,Lapenta1995,Schweigert1996,Lapenta1999,Winske1999,
Manweiler2000,Winske2000,Hutchinson2002,Hutchinson2003,Hutchinson2005,
Hutchinson2006a,Matyash2006,Smirnov2006,Matyash2007,Miloch2007,Miloch2008,
Miloch2009,Miloch2009a,Hutchinson2011a,Miloch2013}. MC methods are also used to
study ion transport in complex plasmas \cite{Sun2000}. The
interaction between multiple microparticles is taken into account with a
particle-particle-particle-mesh scheme. Microparticles are usually
treated as stationary \cite{Block2012}. Some authors have also
performed a molecular dynamics simulation of the plasma flow around
stationary microparticles
\cite{Maiorov2000,Olevanov2003,Vladimirov2003a}. PIC methods can also
be used to study the whole discharge with stationary dust
\cite{Boeuf1992,Chutov2003}. If the PIC is coupled to microparticle
transport modules \cite{Belenguer1992,Lapenta1997,Schweigert2008},
dust distribution functions can be derived. 

The third approach to modeling complex plasmas is to use
molecular-dynamics (MD)/Langevin dynamics simulations of the microparticles. The
plasma is usually taken into account analytically via the microparticle
interaction potential \citetext{e.g.,
  \citealp{Khodataev1998,Winske1999,Samsonov1999a,
Ohta2000,Quinn2000,Vaulina2000,Vaulina2001,Goedheer2003,Jiang2006,Klumov2006,Liu2006a,Liu2007a,Block2008, Kaehlert2008,Ott2008,Jiang2010,Wysocki2010,Couedel2011,Jiang2011,Reichstein2011}}. A
PIC analysis can determine the interaction potential more accurately
and then serve as input for the MD simulation \citetext{e.g., \citealp{Schweigert2002,Jiang2009a}}. Other plasma effects can be taken into
account analytically as well, for instance a fluctuating particle
charge \cite{Vaulina1999}. \citet{Schweigert1998a} use an MD
simulation and include the effect of ion space charges by analytically
including point charges in place of the ion space charges below the
microparticles, as proposed in \cite{Schweigert1996}. 

Hybrid approaches combine two or more of the techniques described
above. For instance, \citet{Matyash2007,Matyash2010} use a P$^3$M model which
combines PIC and MD approaches to study charging of microparticle
strings. If nano- and microparticle growth is to be modeled, often, sectional
models are used. These models can be coupled to fluid models of the 
plasma \citetext{see \citealp{Agarwal2012} and references 
therein}. \citet{Melandso1996a,Melandso1997} use a hybrid approach in which they
represent the plasma as a fluid and the microparticles as diffuse
objects and study particle arrangement under the influence of the ion
drag force. \citet{Yu2006} combine a 2D sheath fluid model and a 3D
microparticle transport model to study the formation of plasma
crystals. \citet{Rome2012} couple a fluid description of electrons to
a kinetic description of dust.

Several authors in the group of M. Kushner have developed a series of
linked computer models to investigate the trapping of microparticles
in radio-frequency (rf) discharges
\cite{Choi1994,Hwang1997,Hwang1998,Vyas2002,Vyas2005}: A
Monte-Carlo-fluid hybrid simulates the plasma, a PIC simulation determines microparticle charging and
ion-microparticle momentum transfer cross section, and a plasma
chemistry Monte Carlo simulation provides ion flux. This is coupled to
a microparticle transport module that calculates microparticle
trajectories. The authors were able to
identify several regions in which microparticle trapping
occurred. In the later models feedback to the plasma was included to investigate
particle transport and Coulomb crystallization \cite{Vyas2002}.

The model we present here follows a similar concept as
\cite{Choi1994,Hwang1997,Hwang1998,Vyas2002,Vyas2005}: We perform a
hybrid fluid/analytical simulation of a radio-frequency plasma and couple this to a
molecular dynamics simulation of the microparticles. In contrast to
the work described above, which was mainly aimed at finding the trapping
positions and studying plasma crystal formation, we reproduce
dynamical collaborative effects in complex plasmas and 
compare with experimental results. 

Both the capacitely-coupled plasma (CCP) model and the MD model are
two-dimensional. The CCP model consists of a plasma sheath, which is
solved analytically, and the plasma bulk, which is subject to the
fluid simulation \cite{Kawamura2011}. The sheath width is fixed during the run of one
simulation. This means that effects that depend on a changing sheath
width, such as melting a plasma crystal by increasing the gas pressure \cite{Khrapak2012b}, cannot
be modelled accurately without extending the model. Also, we do not
include the feedback from the microparticles to the plasma in this
version of the model. This limits the applicability of the model to
low microparticle densities where the Havnes parameter $\mathcal{H}$
\cite{Havnes1984} is less than unity: 
\begin{equation}
  \mathcal{H} = n_d Z_d / n_e \ll 1.
\end{equation}
Here, $n_d$ and $n_e$ designate the number densities of the
microparticles and electrons and $Z_d$ the number of electrons on the
microparticles. 

\section{Plasma model}

\begin{figure}
  \centering
  \includegraphics[width=0.5\linewidth]{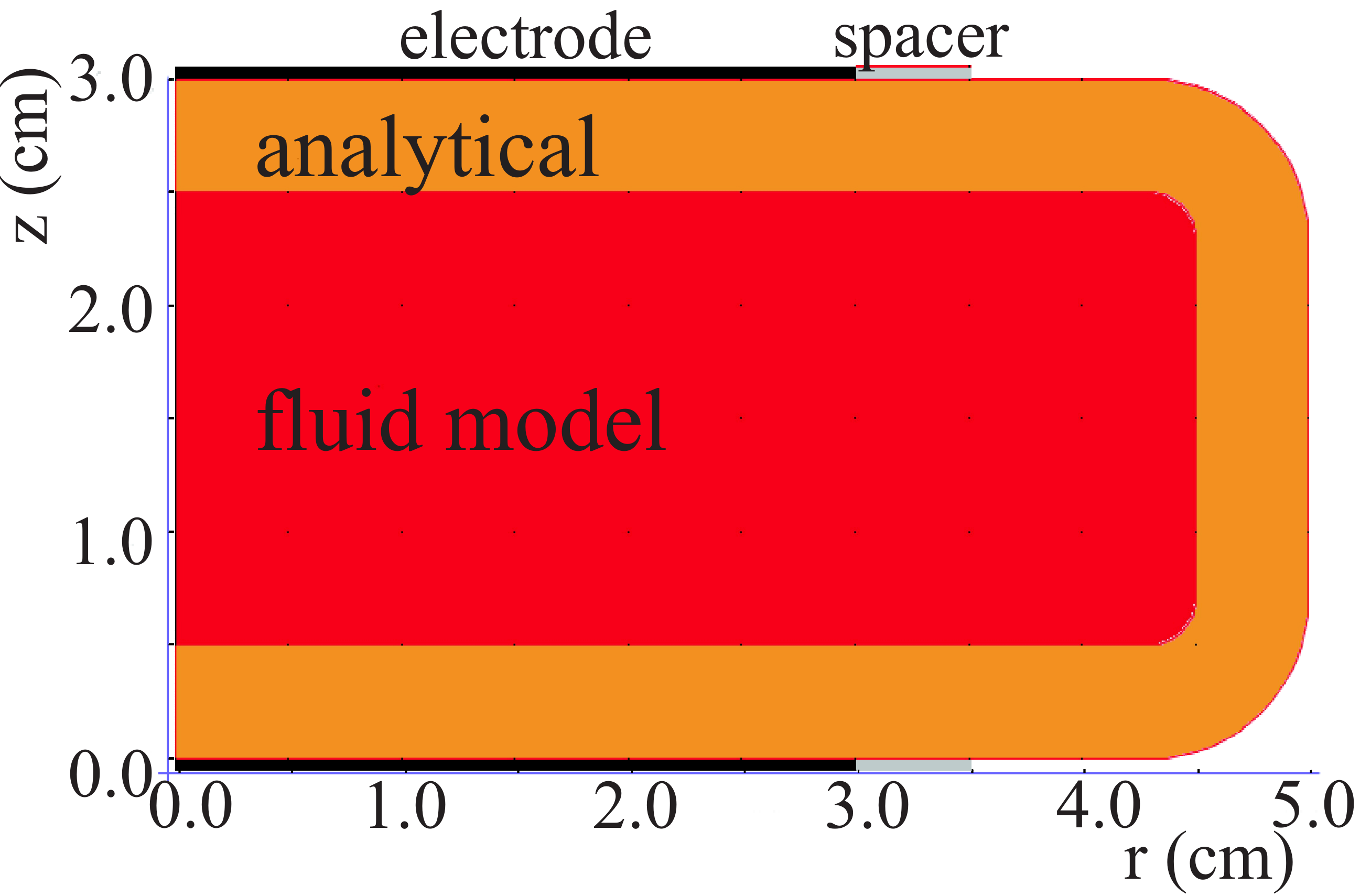}
  \caption{\label{fig:geometry}(color online) Modelled geometry, based
    on the PK-3 Plus chamber \cite{Thomas2008}. Two
    parallel electrodes of radius \SI{3}{\centi \meter} are separated
    by a distance of \SI{3}{\centi \meter}. The electrodes are
    surrounded by spacers of width \SI{5}{\milli \meter}. The nominal sheath
    width is set to \SI{5}{\milli \meter}. The sheaths are solved
    analytically, while the bulk plasma is treated with a fluid
    model. We only model the right half of the plasma chamber midplane
  and assume cylindrical symmetry.}
\end{figure}

Our capacitively-coupled plasma (CCP) model is a modified version of the hybrid fluid-analytical
simulation of inductive/capacitive discharges by \citet{Kawamura2011},
which itself is an extension of the plasma model by 
\citet{Hsu2006} and the analytical sheath model by \citet{Lee2008}. 
We base our model on the geometry of the PK-3 Plus chamber, which
is a parallel-plate, capacitively coupled plasma chamber
\cite{Thomas2008}. Figure~\ref{fig:geometry} shows a sketch of the geometry. The electrodes are separated by \SI{3}{\centi
  \meter} and have a diameter of
$\SI{6}{\centi \meter}$. We only model the right half of the experimental
chamber, as it is approximately cylindrically symmetric. The electrodes are surrounded by a dielectric spacer and a grounded
guard ring. In the physical experiment, the spacer is very small, but
in the simulation, we use a spacer of width \SI{5}{\milli
  \meter}. The reason for this choice is that we have to integrate over the
spacer to obtain some output values, such as the radio-frequency
voltage, and the error in the integration becomes large when small spacers are used. A wider spacer does not
significantly modify the results in the plasma bulk. 

The bulk of the plasma is modeled as a quasi-neutral fluid. Assuming
quasi-neutrality significantly speeds up the simulation compared to
the alternative, solving the Poisson equation. The sheaths are solved
analytically. We fix the nominal sheath width
at \SI{5}{\milli \meter} and take into account the physically varying sheath
width by varying the dielectric constant. All inductive coupling that
is present in \cite{Kawamura2011} is removed from the model. We also
do not model the gas flow or the temperature distribution of the
neutrals and ions in order to make the model faster.

\begin{figure}
  \centering
  \includegraphics[width=0.75\linewidth]{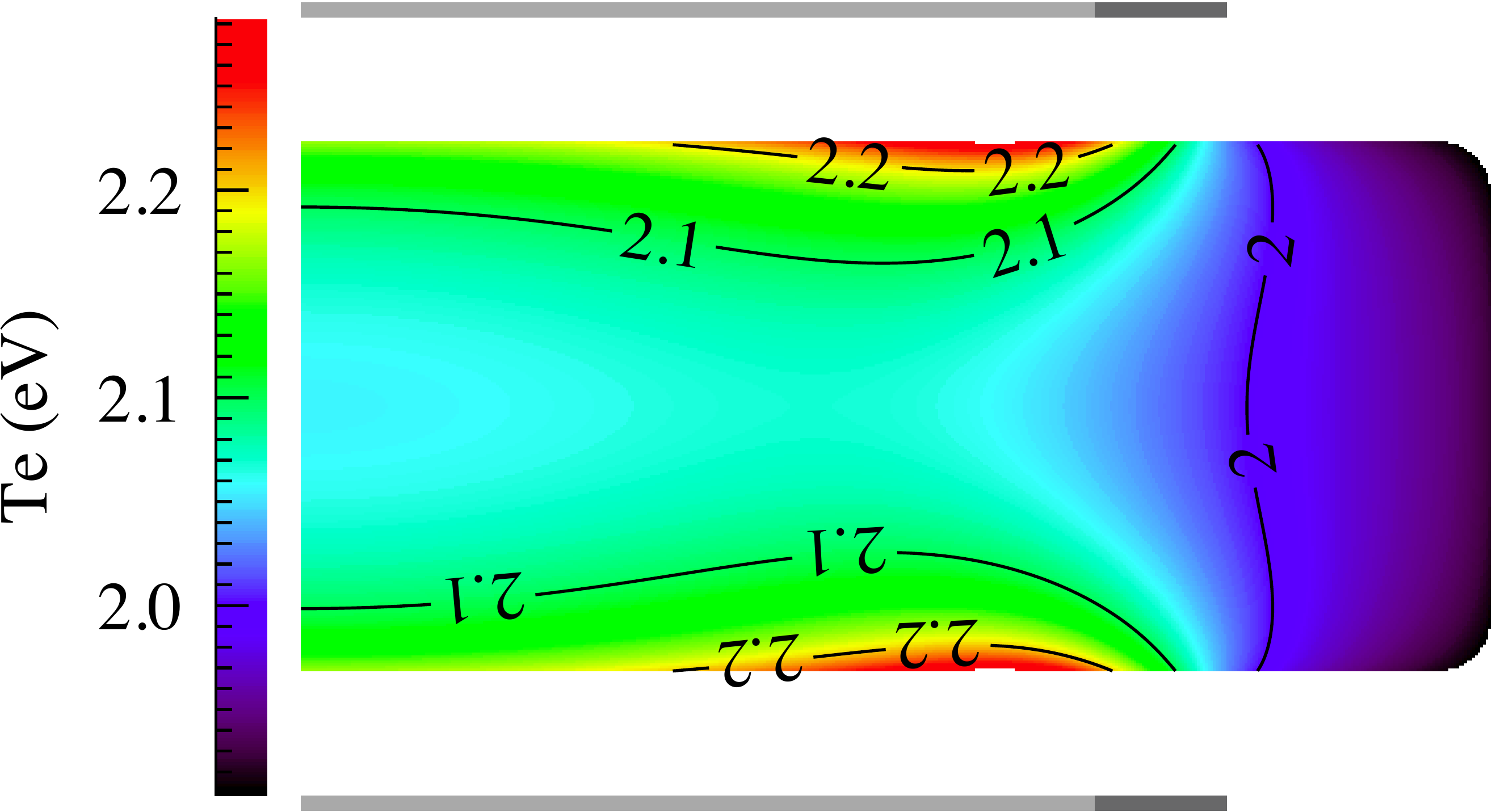}
 \caption{\label{fig:TeDistr}(color online) Electron temperature $T_e$
   in the plasma bulk at a gas pressure of \SI{20}{\pascal} and an input current
    of \SI{20}{\milli \ampere}. The gray lines in the top and bottom
    indicate the position of the electrodes and spacers.} 
\end{figure}

\begin{figure}
  \includegraphics[width=\linewidth]{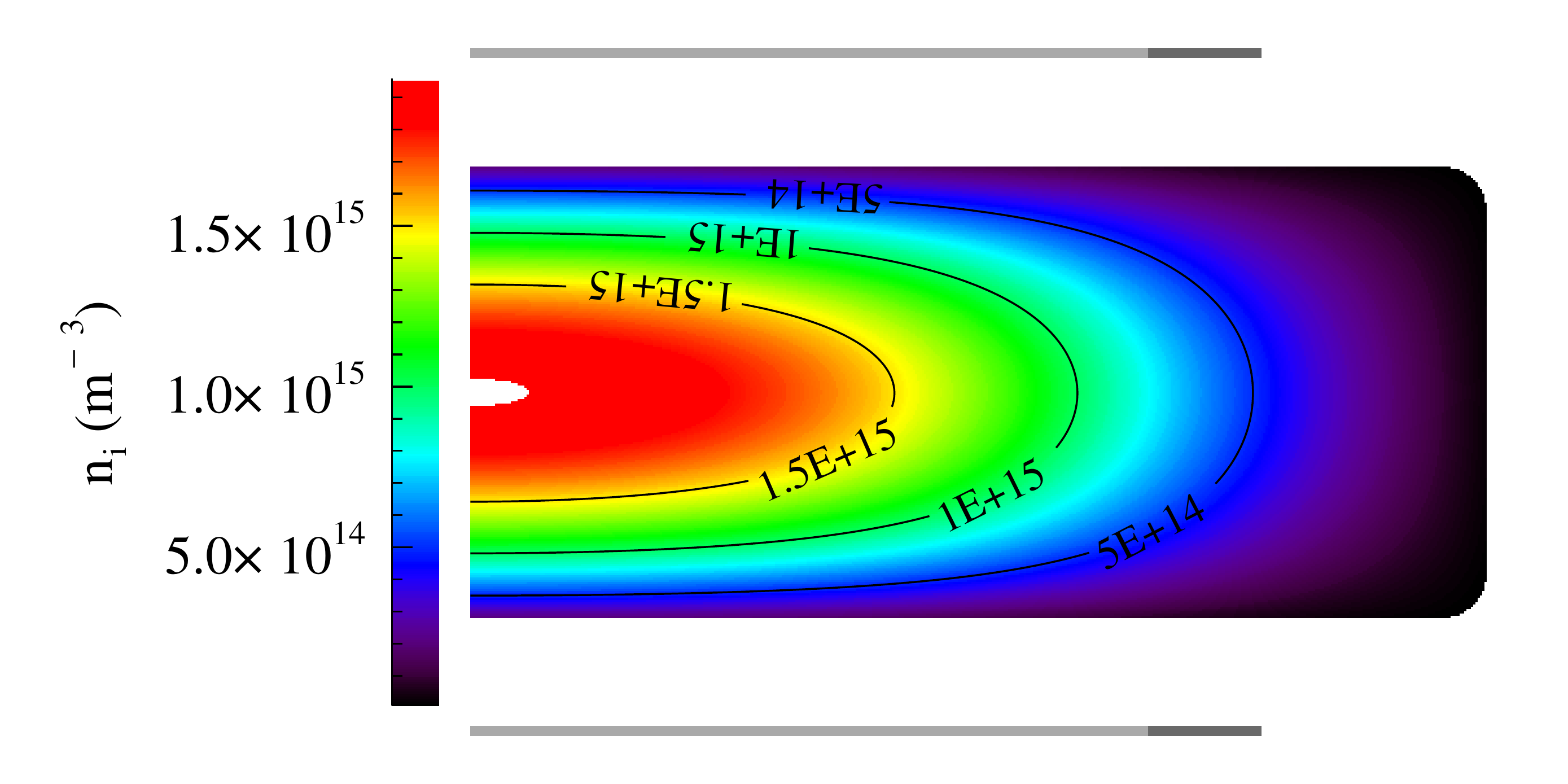}
  \caption{\label{fig:niDistr}(color online) Distribution of ion
    density $n_i$ (and, via quasi-neutrality, electron density) in the plasma bulk at a
    pressure of \SI{20}{\pascal} and an input current of
    \SI{20}{\milli \ampere}. The gray lines in the top and bottom
    indicate the position of the electrodes and spacers.}
\end{figure}

The model uses the gas pressure $p$, the ion temperature $T_i$ and the
input current $I$ as input parameters. We always chose $T_i =
\SI{300}{\kelvin}$ and model argon as buffer gas. Without taking into account the influence of the
microparticles on the plasma, quasi-neutrality causes the electron and ion densities to
always be equal, $n_i = n_e$. The ion velocities are subject to the boundary
condition that they reach Bohm velocity $u_B$ at the sheath edge, with
$u_B = \sqrt{k_B T_e / m_i}$,
where $T_e$ indicates the electron temperature, $k_B$ Boltzmann's
constant, and $m_i$ the ion mass. The simulation results in the plasma
densities, ion flux, electron temperature, ambipolar and rf electric
fields, ionization rate, etc. 

Even though we attempt to make the model as accurate as possible,
taking into account various power deposition and energy loss
mechanisms \cite{Kawamura2011}, we do not believe that this model is a quantitatively absolutely
accurate description of the PK-3 Plus chamber. For instance, the
currents used in the model are generally higher than those determined
experimentally. Nevertheless, the model is qualitatively sufficient to
reproduce many of the phenomena observed in the experiment, as we will
discuss later. Before getting to the microparticle dynamics, we shall
first discuss typical results of the plasma model.

Figure~\ref{fig:TeDistr} shows the electron temperature distribution
in the plasma bulk at a pressure of \SI{20}{\pascal} and an
input current of \SI{20}{\milli \ampere}. It can be seen that the
electron temperature is highest near the electrodes, with a
maximum close to the edge of the electrodes. The difference in
electron temperature near the electrodes and in the center of the
discharge is about 10\%. In contrast to the electron temperature, the
electron and ion densities peak in the center of the 
discharge, as shown in Fig.~\ref{fig:niDistr}. 

\begin{figure}
  \includegraphics[width=\linewidth]{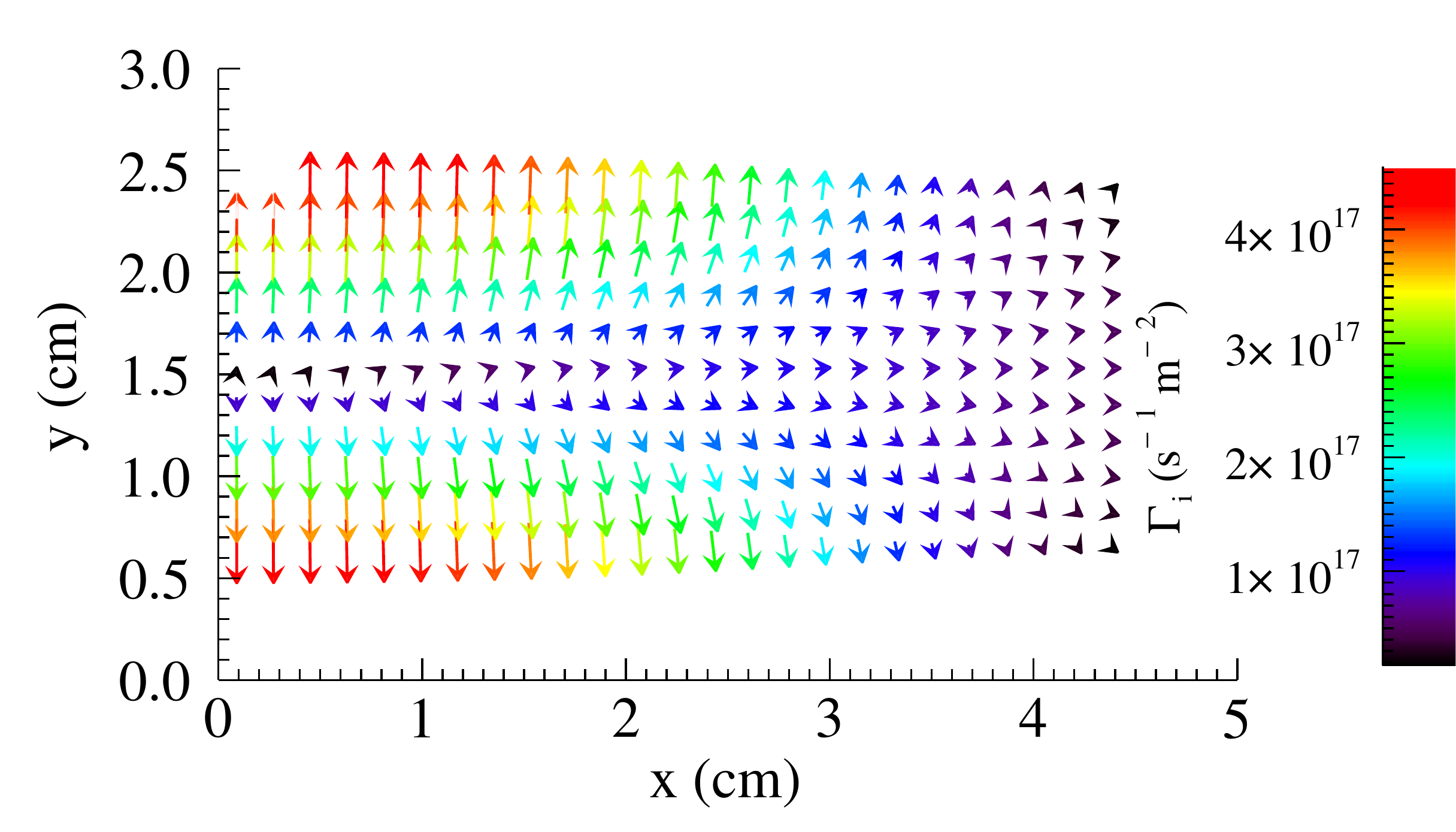}
  \caption{\label{fig:Gi}(color online) Ion flux $\Gamma_i$ in the plasma bulk as
    a function of position for a pressure $p = \SI{20}{\pascal}$ and
    an input current $I = \SI{20}{\milli \ampere}$.}
\end{figure}

\begin{figure}
  \includegraphics[width=\linewidth]{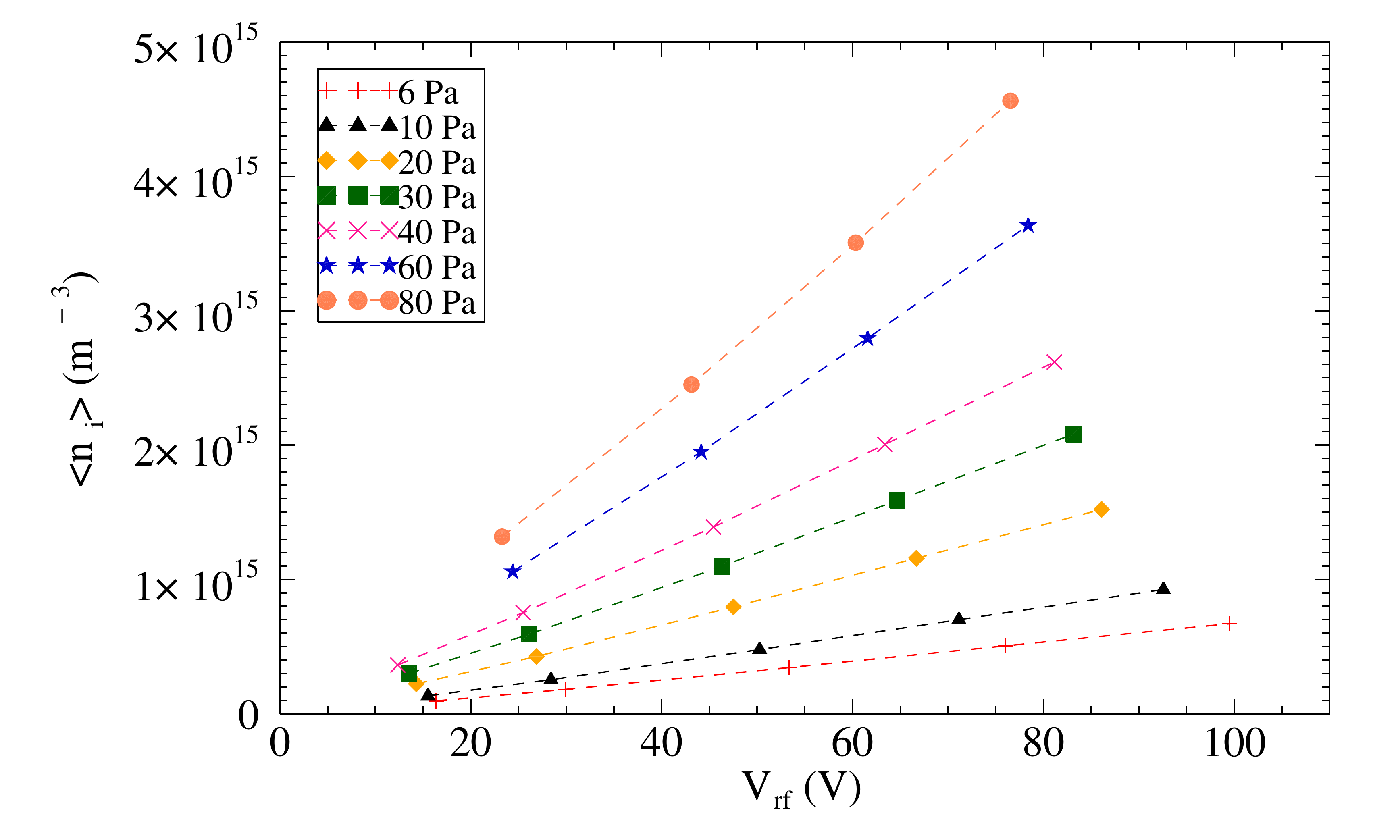}
  \caption{\label{fig:nimean}(color online) Mean ion density as a function of
    rf-voltage for different pressures. The lines are guides to the
    eye that connect series of constant pressure. The input currents
    used for each pressure were $I = \SI{5}{\milli \ampere}$,
    $\SI{10}{\milli \ampere}$, $\SI{20}{\milli \ampere}$,
    $\SI{30}{\milli \ampere}$, and $\SI{40}{\milli
      \ampere}$. For the two highest pressures and lowest input current, the voltage falls
    below \SI{10}{\volt}, and the plasma model is no longer
    applicable. That is why we omit these data in the plot. The case
    shown in Fig.~\ref{fig:TeDistr} and \ref{fig:niDistr} corresponds to the voltage $V_\text{rf} = \SI{47.5}{\volt}$ and the pressure $p =
    \SI{20}{\pascal}$. In general, the ion density increases with
    pressure and with voltage.}
\end{figure}

The ions flow from the center of the plasma bulk outwards. They reach
Bohm velocity at the sheath edge. The distribution of the ion flux is
shown in Fig.~\ref{fig:Gi}. 

The ion and electron densities depend on the pressure and the input
current/voltage, as can be seen in Fig.~\ref{fig:nimean}. They rise
linearly with voltage. The mean density rises with pressure. Simultaneously, the ionization degree
falls (not shown in the figure). 

\begin{figure}
  \includegraphics[width=\linewidth]{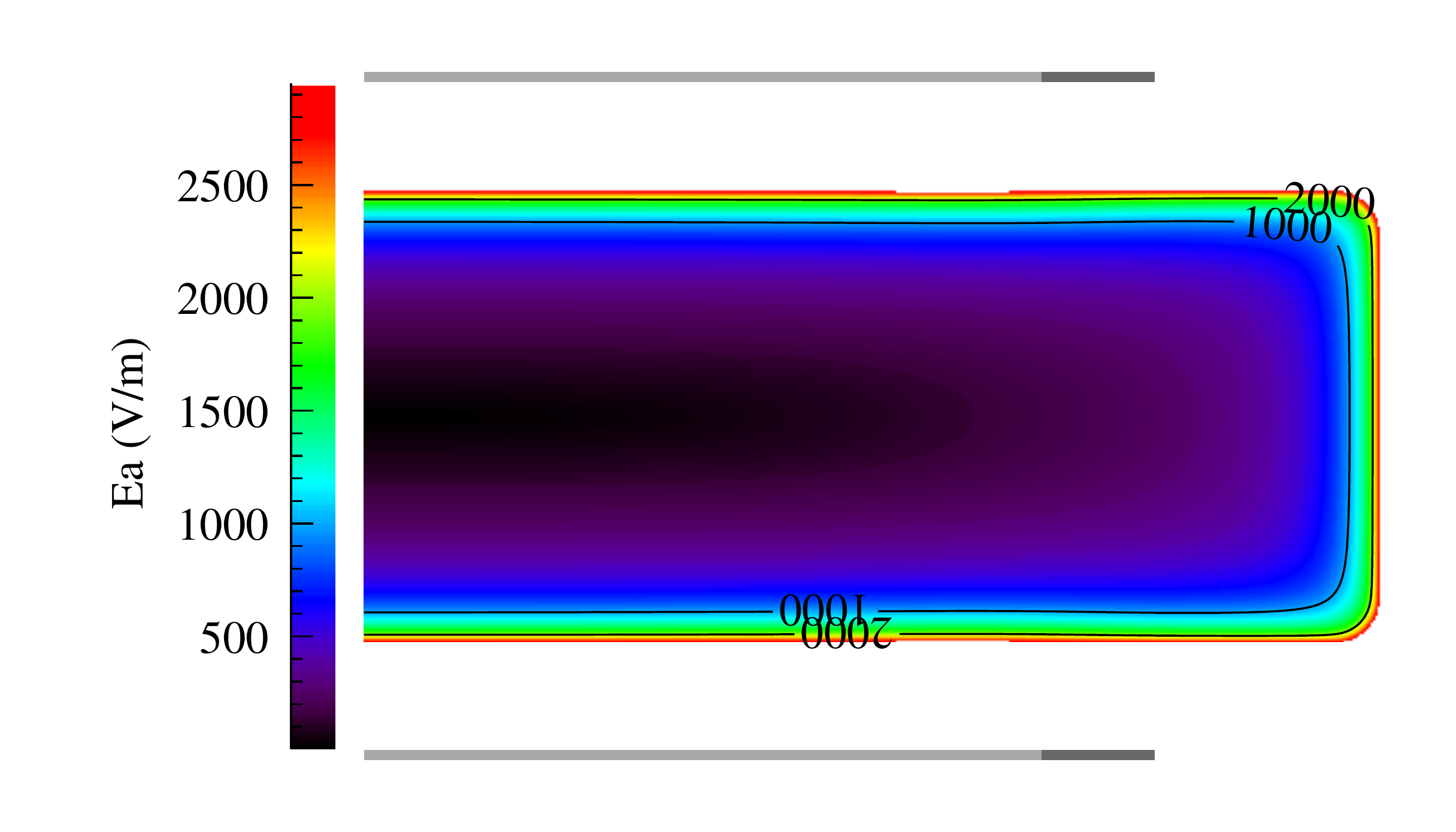}
  \caption{\label{fig:Ea}(color online) Magnitude of the ambipolar
    electric field $Ea$ in the plasma bulk at a
    pressure of \SI{20}{\pascal} and an input current of
    \SI{20}{\milli \ampere}. The gray lines in the top and bottom
    indicate the position of the electrodes and spacers.}
\end{figure}

The densities of the plasma particles, the electron temperature and
the ion flux have a profound impact on the dynamics of microparticles
embedded in the plasma. The microparticles are confined in the bulk of
the discharge by the ambipolar electric field that build up in and
near the sheaths. Figure~\ref{fig:Ea} shows the magnitude of the
ambipolar electric field in the plasma bulk at a pressure of
\SI{20}{\pascal} and an input current of \SI{20}{\milli \ampere}. The
field is strongest close to the sheaths and falls in the center of the
chamber. 

\section{\label{sec:md}Molecular dynamics model of the microparticles}

We use the freely available molecular dynamics (MD) code LAMMPS
\cite{Plimpton1995}. The microparticles are modelled as charged point
particles that interact via a screened Coulomb potential $\Phi$
\begin{equation}
  \Phi = \frac{q_1 q_2}{4 \pi \epsilon_0 r}  \exp(-r/\lambda_D), 
\end{equation}
where $q_1$ and $q_2$ are the charges of the microparticles, $r$ is
the distance between them, $\epsilon_0$ signifies the electric permittivity of the
vacuum, and $\lambda_D$ is the 
Debye length. We use the Debye length calculated from the mean ion
density resulting from the plasma simulation. In future refinements of
the code, a local Debye length could be used, but this is not
implemented yet. It is possible to use a charge of the microparticles
that depends on the position - however, in the simulations presented
in this paper, all microparticles carry the same charge. Typically, we
estimate the average microparticle charge $q$ with
\citet{Matsoukas1995}'s 
approximation
\begin{equation}
  q \approx C \frac{4 \pi \epsilon_0 r_d k_B T_e}{e^2} \ln
  \frac{n_i}{n_e} \left( \frac{m_e T_e}{m_i T_i} \right) ^{1/2}.
  \label{eq:Matsoukas}
\end{equation}
As usual, $r_d$ is the radius of the microparticles, $k_B$ Boltzmann's
constant, $e$ the electron charge,
$n_i$ and $n_e$ the ion and electron densities, $m_i$ and $m_e$ their
masses, and $T_i$ and $T_e$ their temperatures. For a typical argon
plasma, the constant C is approximately $C \approx 0.73$ \cite{Matsoukas1995}.

Unless otherwise noted, we calculate the microparticle charge with Eq.~\ref{eq:Matsoukas} in the
beginning of the simulation as a function of the mean plasma
parameters. As alternative for specific simulations, we set the charge to a specific, user selected value.

The microparticles are embedded in a gas with a given
temperature and interact with the gas. Firstly, there is
friction with the gas when the microparticles are moving with a
velocity $\bs{v}$ with respect to the gas. This results in the force
\begin{equation}
  \bs{F_{Ep}} = - m_d \gamma \bs{v},
\end{equation}
where $m_d$ is the mass of the microparticles, and $\gamma$ is the
coupling constant. We use \citet{Epstein1924} 's well-known formula to calculate
$\gamma$, using the 
coefficient $\delta = 1.48$ as determined experimentally for complex plasmas
\cite{Konopka2000}:
\begin{equation}
  \gamma = \delta \frac{n_n m_n u_n}{\rho_d r_d}.
\end{equation}
Here, $n_n$, $m_n$, and $u_n$ signify the number density, mass, and
thermal velocity of the neutral particles, respectively. The
microparticle mass density and radius are given by $\rho_d$ and
$r_d$. 

Secondly, the gas not only reduces the microparticle velocity, it also
transfers energy to the microparticles via heating. This is
modeled by random kicks to the microparticles that bring the
microparticles to the same temperature as the background gas. The
resulting force $\bs{F_r}$ is proportional to
\begin{equation}
  F_r \propto \sqrt{\frac{k_B T_n m_d \gamma}{dt}},
\end{equation}
with $dt$ being the time step of the simulation and $T_n$ the gas
temperature. The force is applied on a per-particle basis. The
direction and magnitude is randomized using uniform random numbers
\cite{LammpsDoc}. 

We have modified the LAMMPS source code to include the influence of
the plasma. For this purpose, the ion and electron densities, electron
temperature, ion velocities and ambipolar electric fields are written to an ascii file by Matlab
using the information from COMSOL. Our C++
routine reads this file and interpolates the information to the positions of the
microparticles. 

The microparticles with charge $q_d$ react directly to the ambipolar electric field $\bs{E_a}$ that
is output from the plasma simulation. This results in the electric force
\begin{equation}
  \bs{F_e} = q_d \bs{E_a}.
\end{equation}

We calculate the ion drag force $\bs{F_{id}}$ using the same approach as
\citet{Goedheer2009}, namely by incorporating results from previous studies
by \citet{Khrapak2002}, \citet{Ivlev2004}, and \citet{Hutchinson2006a}:
\begin{equation}
\begin{split}
  \bs{F_{id}} = & \, n_i m_i u_i \bs{v_i} \biggl( \sigma_c(u_i) + \\ 
  &\pi \rho_0
  (u_i)^2 \left[ \Lambda(u_i) + \mathcal{K}\left(
      \frac{\lambda_D(u_i)}{l_{\text{mfp}}}\right)\right] \biggr).
\end{split}
\end{equation}
Here, $n_i$ is the ion number density, $m_i$ the ion mass, $u_i$
the mean ion velocity, $\bs{v_i}$ the ion velocity, $\sigma_c$ the
cross section for ion capture, $\rho_0$ the Coulomb radius, $\Lambda$
the Coulomb logarithm and $\mathcal{K}$ a collisional function to take
into account loss of angular momentum in collisions of ions and
neutrals \cite{Ivlev2004}. The Debye length $\lambda_D$ is the
linearized Debye length based on the ordinary electron Debye length
and the ion Debye length derived from the total ion energy. The ion
mean free path is denoted by $l_{\text{mfp}}$.
The ion capture cross section $\sigma_c$ is a function of the radius of the
microparticles $r_d$ and the Coulomb radius $\rho_0 = Z_d e^2/2 \pi
\epsilon_0 m_i u_i^2$, where $Z_d$ denotes the number of
electron charges on the microparticles, $\epsilon_0$ the vacuum
permittivity, and $e$ the electron charge:

\begin{equation}
  \sigma_c(u_i) = \pi r_d^2 \left(1 + \frac{\rho_0(u_i)}{r_d}\right).
\end{equation}

The Coulomb logarithm $\Lambda$ is used in the calculation of the cross-section for
scattering of ions around the microparticle. Following
\citet{Khrapak2002}, we include ions scattered at a distance larger
than the Debye length, and use
\begin{equation}
  \Lambda(u_i) = \ln \left[ \frac{\rho_0(\tilde{u_i}) + \lambda_D(\tilde{u_i})}{\rho_0
      (\tilde{u_i}) + r_d} \right].
\end{equation}

In the calculation of the Coulomb logarithm, the expression of the mean
ion velocity $u_i$ is modified from its ordinary value,  $u_i^2 = 8
  k_B T_{gas} / \pi m_i + \bs{v_i}^2$, to fit a PIC simulation also
at higher ion flow speeds. We use the correction factor given in
\cite{Hutchinson2006a}, rather than the simpler one from
\cite{Hutchinson2006} used in \cite{Goedheer2009}:
\begin{equation}
\begin{split}
  &\tilde{u_i}^2 =\frac{8 k_B T_{gas}}{\pi m_i} + \bs{v_i}^2 \biggl( 1
  +\\ &\left[ \frac{|\bs{v_i}|/ u_B}{0.6 + 0.05 \ln(m_a) +
      (\lambda_{De}/5r_d)(\sqrt{T_i/T_e}-0.1)} \right]^3 \biggr).
\end{split}
\end{equation}
Here, $m_a$ denotes the atomic mass of the ions, $m_a = 40$ for
Argon. We also assume singly charged ions. The Bohm velocity $u_B$ is
given by $u_B = \sqrt{k_B T_e / m_i}$.

The collisional function $\mathcal{K}$ is given by
\cite{Ivlev2004}
\begin{equation}
\begin{split}
  \mathcal{K}(x) = &x \arctan(x) + \left(\sqrt{\frac{\pi}{2}}-1\right)
    \frac{x^2}{1+x^2} - \\ &\sqrt{\frac{\pi}{2}} \ln (1 + x^2).
\end{split}
\end{equation}

We do not model the distribution of neutral temperature in the chamber
and thus do not take into account the thermophoretic force. Even
though we do have a slight electron temperature gradient (see
Fig.~\ref{fig:TeDistr}), so far we do not take into account the force
due to the electron temperature gradient \cite{Khrapak2013}. 

Finally, after all forces are determined, the positions and velocities of the microparticles are
updated with the Velocity Verlet algorithm \cite{LammpsDoc}. We
usually use time steps of \SI{0.1}{\milli \second} or \SI{1}{\milli
  \second}, depending on the velocity of the microparticles. We use a
cutoff of the potential of \SI{2}{\milli \meter} for very dense
complex plasmas up to \SI{1}{\centi \meter} for very dilute complex
plasmas.

\section{Collective particle effects}

Complex plasmas display a wealth of collective effects. In the
following, we shall discuss some effects that our model is able to
reproduce. Note that our model is two-dimensional (2D), whereas experiments
with complex plasmas under microgravity are always
three-dimensional (3D). This leads to a limited comparability, for
instance, crystallization occurs for lower coupling parameters in 2D
than 3D \cite{Bonitz2008}. Correspondingly, the particles in our
simulation are often in crystalline state, whereas they are still
fluid in 3D experiments. 
 
All examples shown model an argon plasma containing spherical
melamine-formaldehyde particles with a mass density of $\rho_d =
\SI{1510}{\kilo \gram \per \cubic \meter}$.

\subsection{\label{sec:confinement}Confinement and void formation}

\begin{figure*}
   \includegraphics[width=0.49\linewidth]{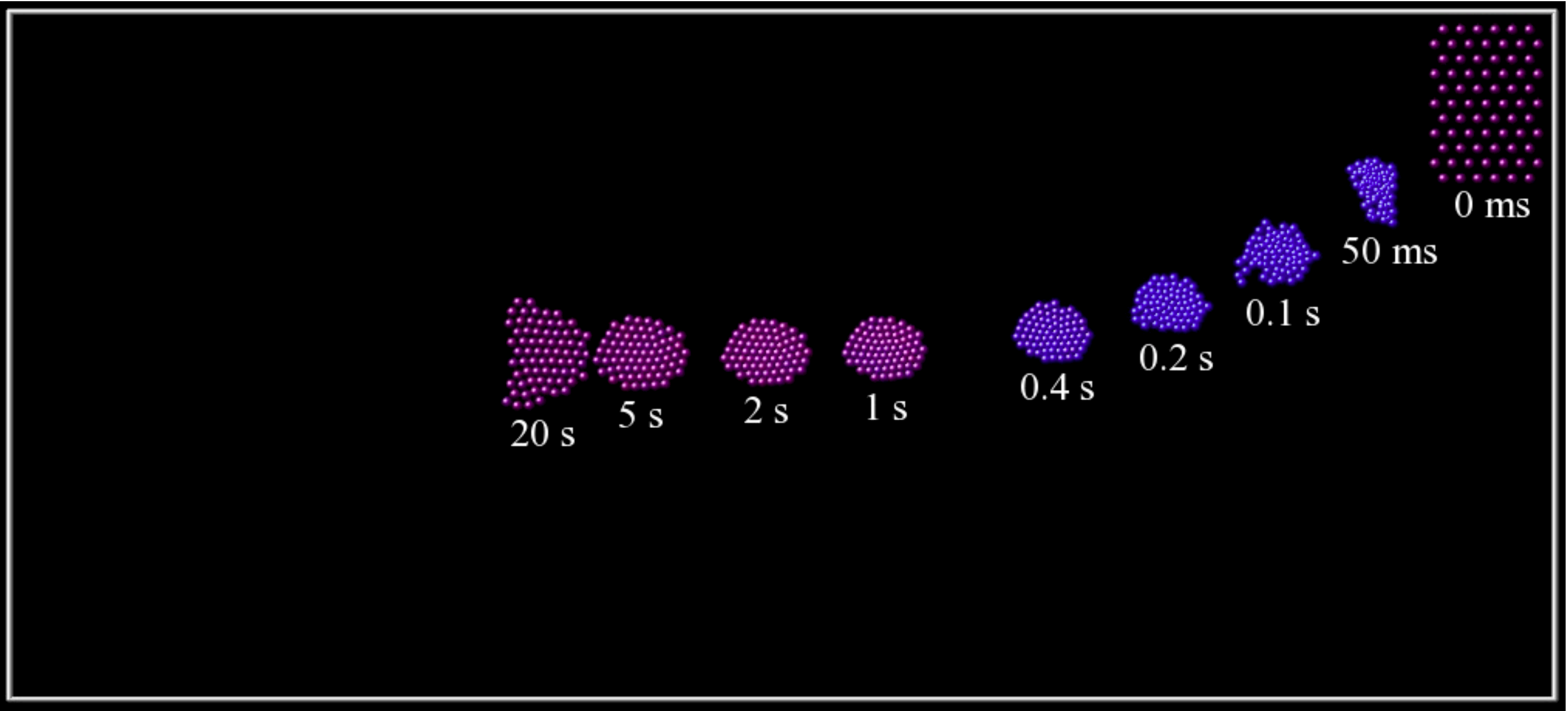}
  \includegraphics[width=0.49\linewidth]{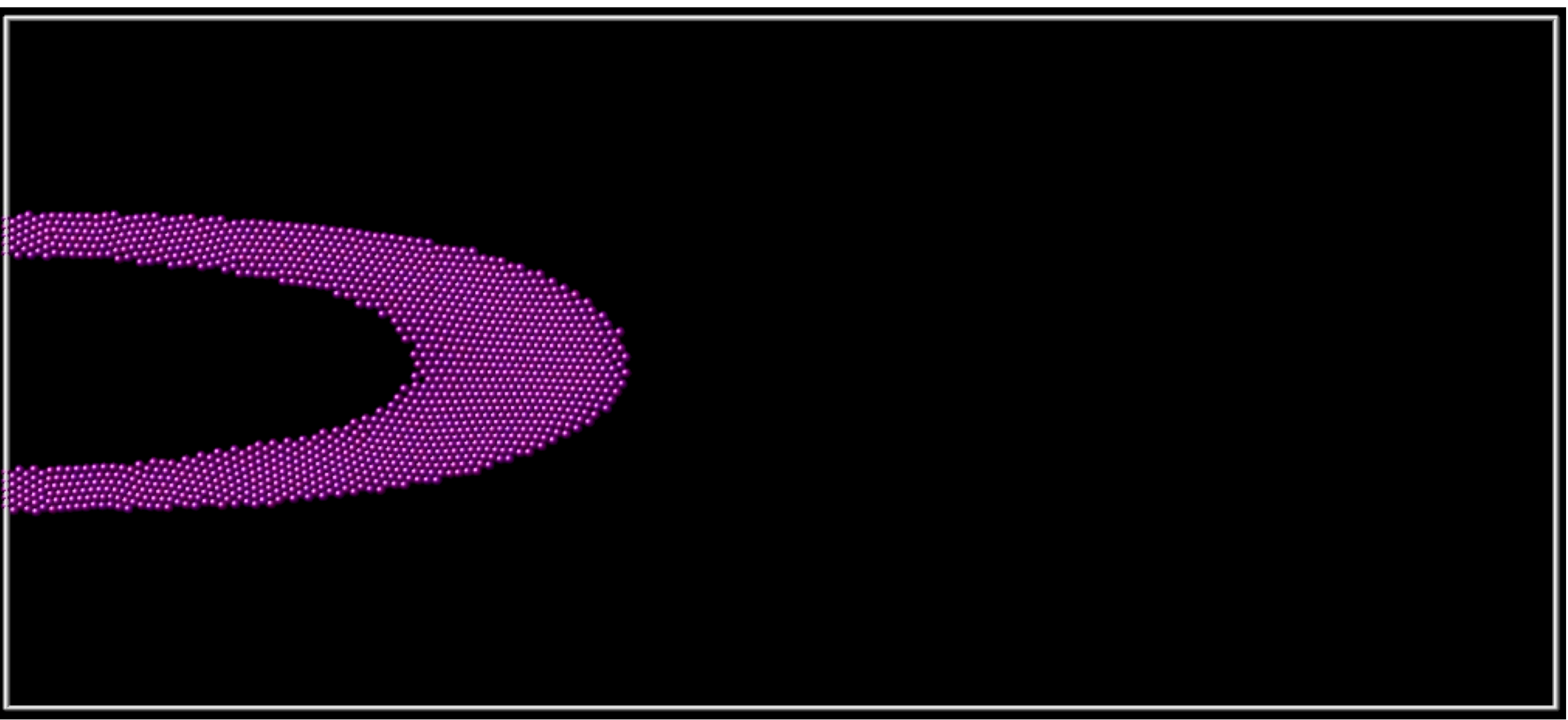}
  \caption{\label{fig:injection}(color online) Formation process of a particle cloud with
    (left) a superposition of nine frames during the injection process and
    (right) the equilibrium particle cloud after 20
    injections and a time span of \SI{3.5}{\minute} after the last
    injection. The gas pressure is \SI{20}{\pascal}, the input current
    \SI{25}{\milli \ampere}. The gray box in the images indicates the
    dimensions of the simulated plasma bulk - one half of the vertical midplane of the plasma chamber without sheaths, with dimensions $\SI[product-units = power]{4.5
      x 2}{\centi \meter}$. The center line of the plasma
    chamber corresponds to the left gray line at the edge of the field of view. The color of the microparticles indicates the vertical velocity $v_z$, ranging from blue/light gray for \SI{-2}{\milli
      \meter \per \second} to red (not reached in the figure) for
    \SI{+2}{\milli \meter \per \second}. Purple/dark gray color indicates no movement in $z$
    direction, as for the time $t = 0$~ms in the left panel.}  
\end{figure*}

In experiments, microparticles are often injected into the plasma
chamber with a dispenser that is mounted in the electrode \cite{Nefedov2003} or
near it \cite{Thomas2008,Land2009}, or from the side \cite{Klindworth2006}. The microparticles
are charged very quickly compared to the time they need to fall from
the dispenser to the lower electrode. Once charged, they are
subject to the electric confinement and ion drag forces. Close to the
sides of the chamber, the confinement force dominates and pushes the
microparticles into the chamber, where they form a cloud. Near the center, the ion drag force dominates over the confinement
force and pushes the microparticles towards the edges of the
plasma. This leads to the formation of the particle-free central
void. 

At some distance from the chamber center, the confinement, interparticle and ion
drag forces balance, and the particle cloud forms. This mechanism has
been modeled before, see for instance \cite{Goedheer2009,Land2010}.

In order to mimic the injection process in the simulation, we place a small number of
particles with no initial velocity near the upper right edge of the
plasma bulk, close to
the sheath. This is shown in the left panel of Fig.~\ref{fig:injection} with the
label ``0~ms''.  Then we let the simulation run its course. The
particles are immediately accelerated towards the center of the
chamber, as can be seen in the subsequent time steps in
Fig.~\ref{fig:injection}. The particle cloud quickly takes a
round shape. This particle droplet then moves
towards the plasma chamber. It is slowed progressively down by Epstein
and ion drag while it
approaches the center, and at some point the particles closer to the
top of the droplet start moving upwards. This is visible in
Fig.~\ref{fig:injection} as a change in the color of the
microparticles. Once the particles approach the chamber center, the
void clearly indents the particle droplet, and the microparticles
start flowing around the void edge to the top and bottom. The
beginning of this process can be seen in the last time step in
Fig.~\ref{fig:injection}. 

We note that the particles move more slowly in the simulation than in
experiments - for similar settings, the particles spread around the
void within a few seconds in experiments. In the simulation, the
particles have just reached the edge of the void after \SI{20}{\second}. The reason
for this is probably that in experiments, the particles are
accelerated more strongly during the injection process: They transverse the
sheath region on their way to the chamber center, where the electric
field is much stronger than in the presheath region. In the
simulation, we treat the sheath analytically, and thus do not allow
particles to enter the sheath. This leads to significantly smaller
accelerations of the microparticles.  

Once the microparticles have reached the void, they spread around
it. The right panel of Fig.~\ref{fig:injection} shows the equilibrium cloud after 20
particles injections such as shown in the left panel of Fig.~\ref{fig:injection}, and 
after the cloud has evolved for 3~minute. More particles are arranged to the
side than above and below the void, as observed in experiments. There
is one layer of particles at the left edge of the simulation
domain -- the center of the plasma chamber. Here, we employ reflective
boundary conditions, and there is no confinement field that would push
the particles away from the edge as at all other edges. The pressure
from the surrounding particles and the boundary lead to
crystallization at this domain boundary. In experiments, it is often
observed that particles crystallize above or below the void, but not
necessarily in vertical strings as is induced by the boundary here. 

In the simulations, the void seems slightly more oval than
in the experiments. This might be due to an imperfect choice of the
sheath width. We will explore this topic in more detail in the
future. In the future, we will also study how a three-dimensional
simulation changes the observed injection process. Experiments,
e.g. on lane formation \cite{Suetterlin2009}, have
shown that particles generally move within the plane in which they
were injected, even before they start to interact with particles that
are already inside the chamber, so we do not expect a large deviation from the behavior
we observe here.

In general, the distribution of particles around the void is stable,
as long as the input parameters are not changed. The positions of
individual particles are not necessarily stable, as we shall see in
the next following.

\subsection{Vortices}

\begin{figure}
  \centering
  \includegraphics[width=0.75\linewidth]{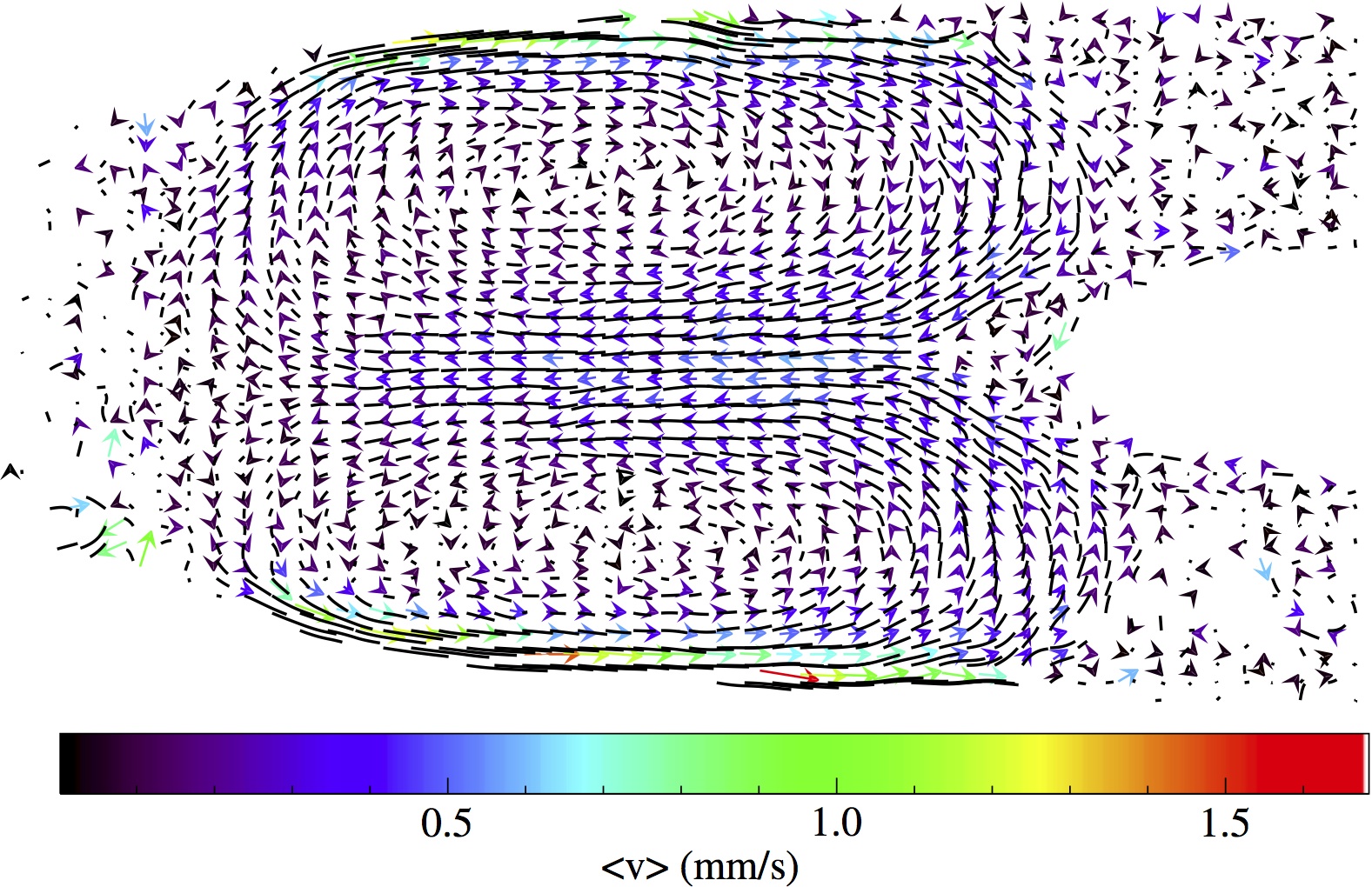}
  \caption{\label{fig:VortexPK3P}(color online) Vortices composed of particles of
    diameter 3.4~$\mu$m in the PK-3 Plus setup under
    microgravity conditions in argon at a pressure of 10~Pa. The
    black lines show the streamlines of the particle movement. The vector field of the mean particle
    velocities is overlaid. The color and length of the vectors indicates the
    total velocity, their orientation the direction of
    movement. Dimensions of the particle cloud: \SI{33 x 16}{\milli \meter}. Data courtesy of the PK-3 Plus
    team \cite{Thomas2008}.}
\end{figure}

Vortices in complex plasmas are common
\cite{Law1998,Morfill1999a,Vladimirov2001,Nefedov2003,Uchida2009}. In the PK-3
Plus setup, they occur at the edges of the cloud and are especially strong at lower
pressures \cite{Morfill1999a}. Figure~\ref{fig:VortexPK3P} shows
overlays of particle streamlines and average velocities recorded at a
pressure of 10~Pa. The particles move along the outer edges of the
cloud towards the cloud center. When they reach the region above or
below the void, they move into the cloud and back outwards. This
direction of the vortices is not always the same, under some
experimental conditions, they can reverse direction (e.g., particles
move towards the void in the center of the cloud and back outwards
along the cloud edge.)  

The mean measured velocity of the particles in the vortices shown in
Fig.~\ref{fig:VortexPK3P} is of the order of 1~mm/s.

\citet{Akdim2003} modeled vortices in complex plasmas using tracer particles
in a fluid simulation. They explain the formation of the vortices as
follows: There is one equilibrium position in the midplane of the
chamber where the confinement force equals the ion drag force. A
single particle will find this position and stay there. As soon as
other particles are present, they displace the particle from the
equilibrium position into a region where the total force is not
conservative. It is there that vortices form. This explanation is confirmed by an MD simulation
\cite{Goedheer2003}. It has been speculated that charge
fluctuations play a role in vortex formation \cite{Vaulina2000} --
however, \citet{Akdim2003} do not find charge differences necessary
for vortex formation.

We confirm this result. Charge fluctuations are not included in our model, and the vortices form nevertheless.
In contrast to \cite{Akdim2003}, in our model, we do not only find one
stable position in the chamber midplane, but a stable circle around
the void. If the neutral gas pressure is low enough, vortices appear
when two or more layers of microparticles are present and their
positions are only slightly removed from the equilibrium line. 

\begin{figure}
  \centering
  \includegraphics[width=0.75\linewidth]{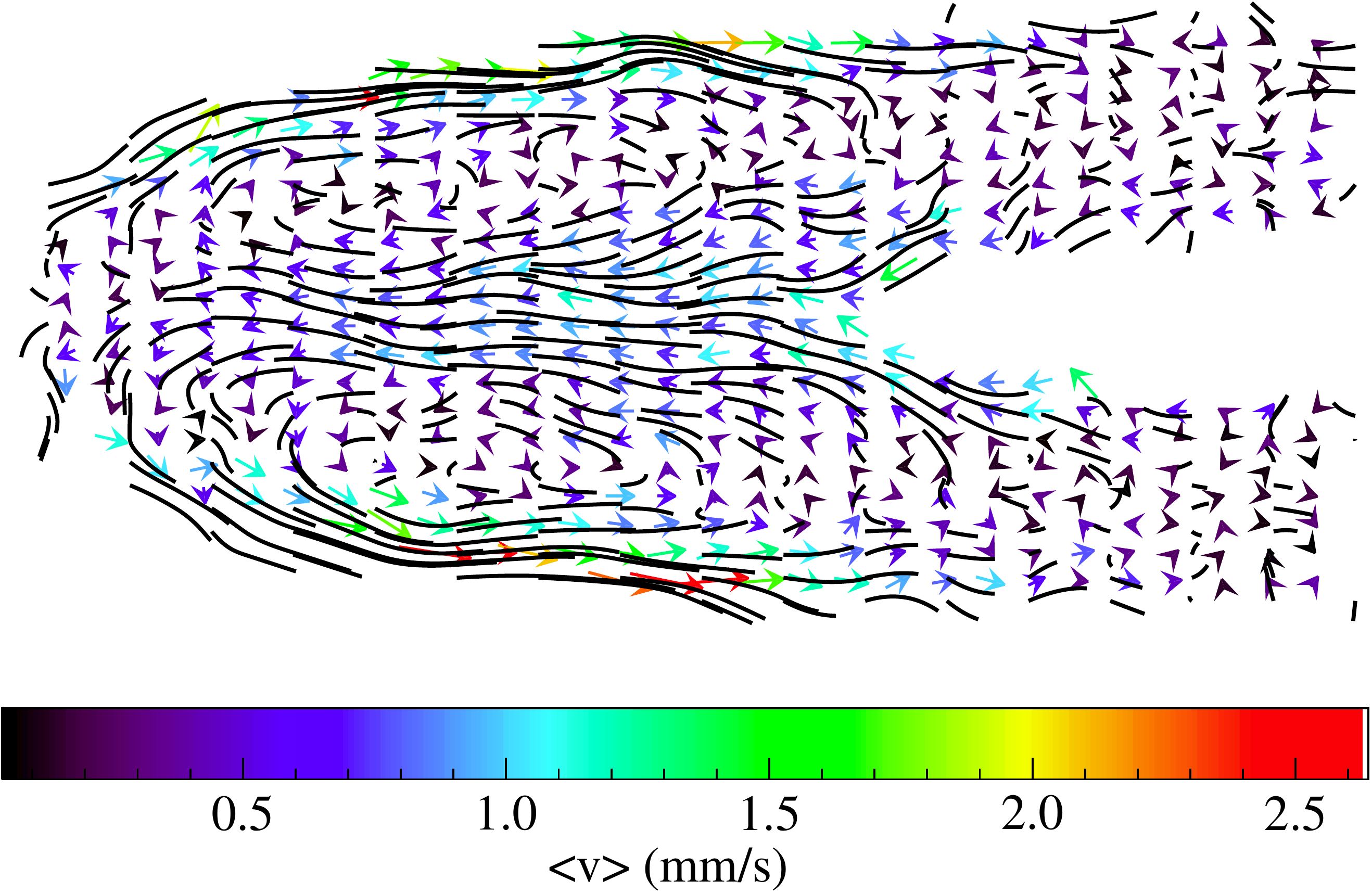}
  \caption{\label{fig:vortex}(color online) Simulated vortex for a pressure of 10~Pa,
    3.4~$\mu$m diameter particles, and an input current of
    16~mA. The particle charge was set to 1/2 of that obtained with
    OML theory. Dimensions of the particle cloud: \SI{23 x
      10}{\milli \meter}. The streamlines of
    the particle movement are shown in black, analogous to Fig.~\ref{fig:VortexPK3P}. Superimposed over this are
    the mean particle velocities. The length and the color of the
    vectors indicates the total velocity, their orientation the
    direction of movement. The image was mirrored to the left side of
    the void to ease comparison with Fig.~\ref{fig:VortexPK3P}.}
\end{figure}

Figure~\ref{fig:vortex} shows particle streamlines and velocities in
the simulation. Vortices are present in which the particles move towards the center of
the cloud along the cloud edges and outwards in the vertical cloud
center. The velocity of the particles depends on the particle
charge. In the figure, we chose a charge of 1/2 of that given by
Eq.~(\ref{eq:Matsoukas}), which results in velocities of up to
\SI{2.6}{\milli \meter \per \second}. 

We also have so far not encountered vortices that rotated in the reversed direction in the simulation. Another
difference to the experimental observations is that the vortices in
our simulations envelop a larger region - they reach almost to the
horizontal center, leaving only a few particles that are in
crystallized state and don't rotate above and below the void. In the
example shown in Fig.~\ref{fig:VortexPK3P}, the void is also
surrounded by a circle of non-rotating particles -- probably smaller
contaminants. In the experiments, there are typically more particles in the regions above
and below the void that do not participate in the rotation. In
experiments, more complicated structures with several vortices along
the horizontal axis and at the outer edges also occur \cite{Nefedov2003}. 

We shall leave the topic of vortices for now and instead complicate
the situation by introducing a second particle type. 

\subsection{Separation by particle size}

When particles of different sizes are injected into a plasma, they
arrange around the void in layers -- the smaller particles closer to
the void than the larger ones. The reason for this is that the
confinement force and the ion drag force have a different dependence
on the particle radius, so that the equilibrium positions of the
different particle sizes vary. 

This has previously been modeled by
\citet{Liu2006a,Liu2007a} via an MD simulation of microparticles in a
quadratic potential. The particles are subject to an outwards pointing
drag force. The authors vary the strength of the drag force and produce
various configurations of microparticles, with the different sizes
mixed, the larger particles on the outside or the inside of the
smaller ones.

\begin{figure}
  \includegraphics[width=\linewidth]{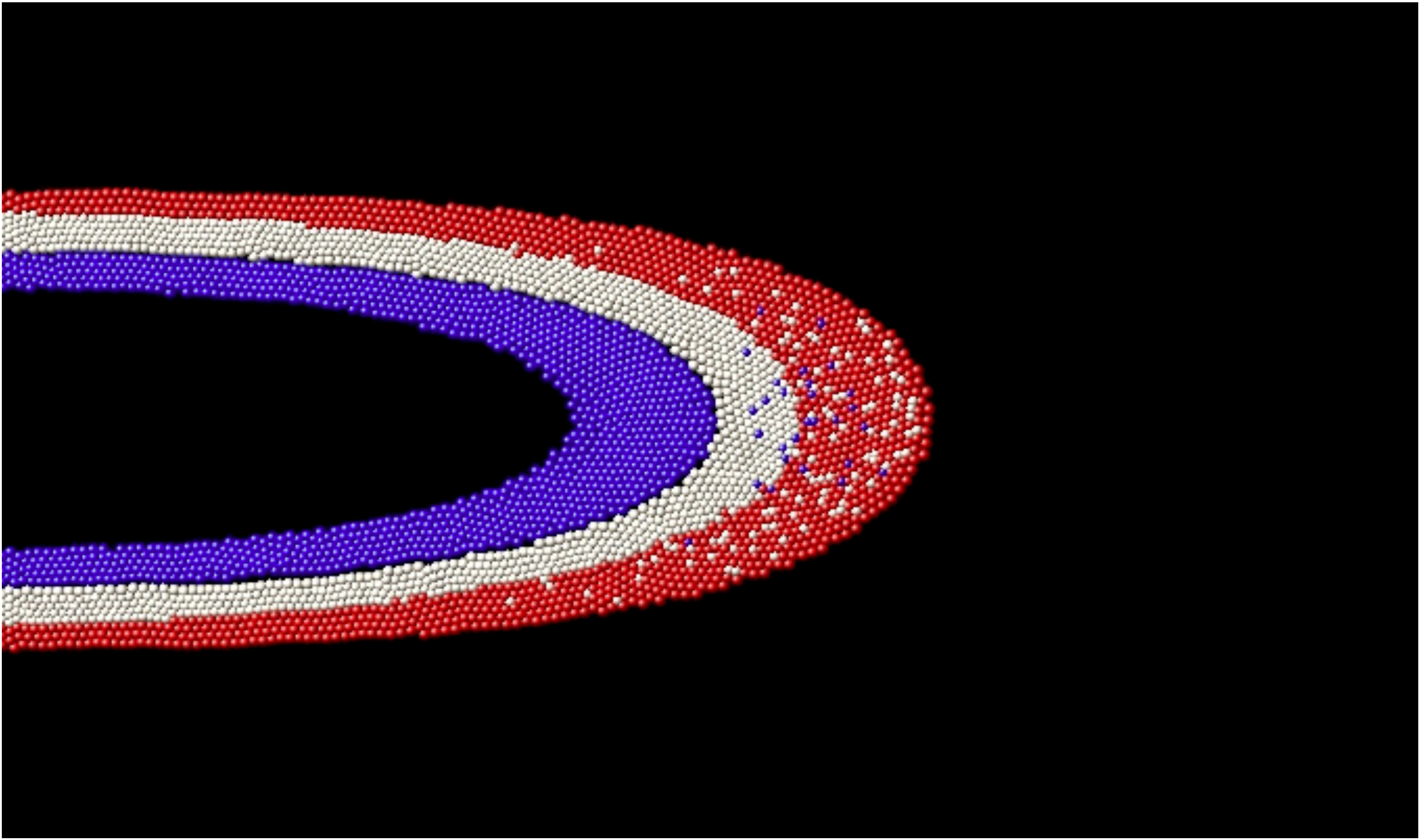}
  \caption{\label{fig:3layers}(color online) Particle cloud made up of three
    different particle sizes: 6.8~$\mu$m (red/light gray), 3.4~$\mu$m (white),
    and 1.55~$\mu$m (blue/dark gray) diameter at a pressure of 20~Pa and a
    current of 30~mA. Field of view: 33.9~mm $\times$ 20.0
   ~mm. The particles were injected in five sets, each composed of a bunch of each particle size. Within
    each set, the largest particles were injected first. After the
    injections, the cloud was left alone to evolve for 3.5~min, during which it sorted by particle size. Even after this
  time, the sorting is not complete -- there are still some smaller
  particles mixed into the larger ones further out.} 
\end{figure}

In our case, we use the expressions for the confinement force and the
ion drag force given in Sec.~\ref{sec:md} and inject particles by placing them at the
edge of the simulation box, as described in
Sec.~\ref{sec:confinement}. The particles automatically arrange in layers as
observed in experiments. Figure~\ref{fig:3layers} shows a particle
cloud composed of microparticles of three sizes that autonomously
arranged in layers. We did not observe homogeneous mixing of the
particle sizes or a reverse of the positions as in \citet{Liu2006a},
in accordance with the experiments. We did, however, observe another
effect that is common in complex plasmas, which we shall discuss next.

\subsection{Lane formation}

Lanes form in complex plasmas when microparticles are injected into a
cloud a larger microparticles. As we have just discussed, the
particles arrange by size -- the smaller particles move to the central
region through the cloud of larger particles. Often, they do so by
forming lanes \cite{Suetterlin2009,Jiang2010,Du2012a}. During this
process, the particles go through three stages
\cite{Du2012a}. Firstly, the ambipolar electric field pushes them
towards the cloud of background particles. Secondly, once they enter
the cloud, they form lanes. The lanes are made up of both small and
large particles. Next, a crossover to the third stage, phase
separation, occurs: The small particles move closer together, forming a drop
inside the cloud of large particles. During the crossover, some large
particles are left over inside the droplet of small particles and form
lanes inside this droplet. Finally, all large particles are expelled, and
a droplet of small particles moves towards the center of cloud.

\begin{figure}
  \centering
  \includegraphics[width=0.5\linewidth]{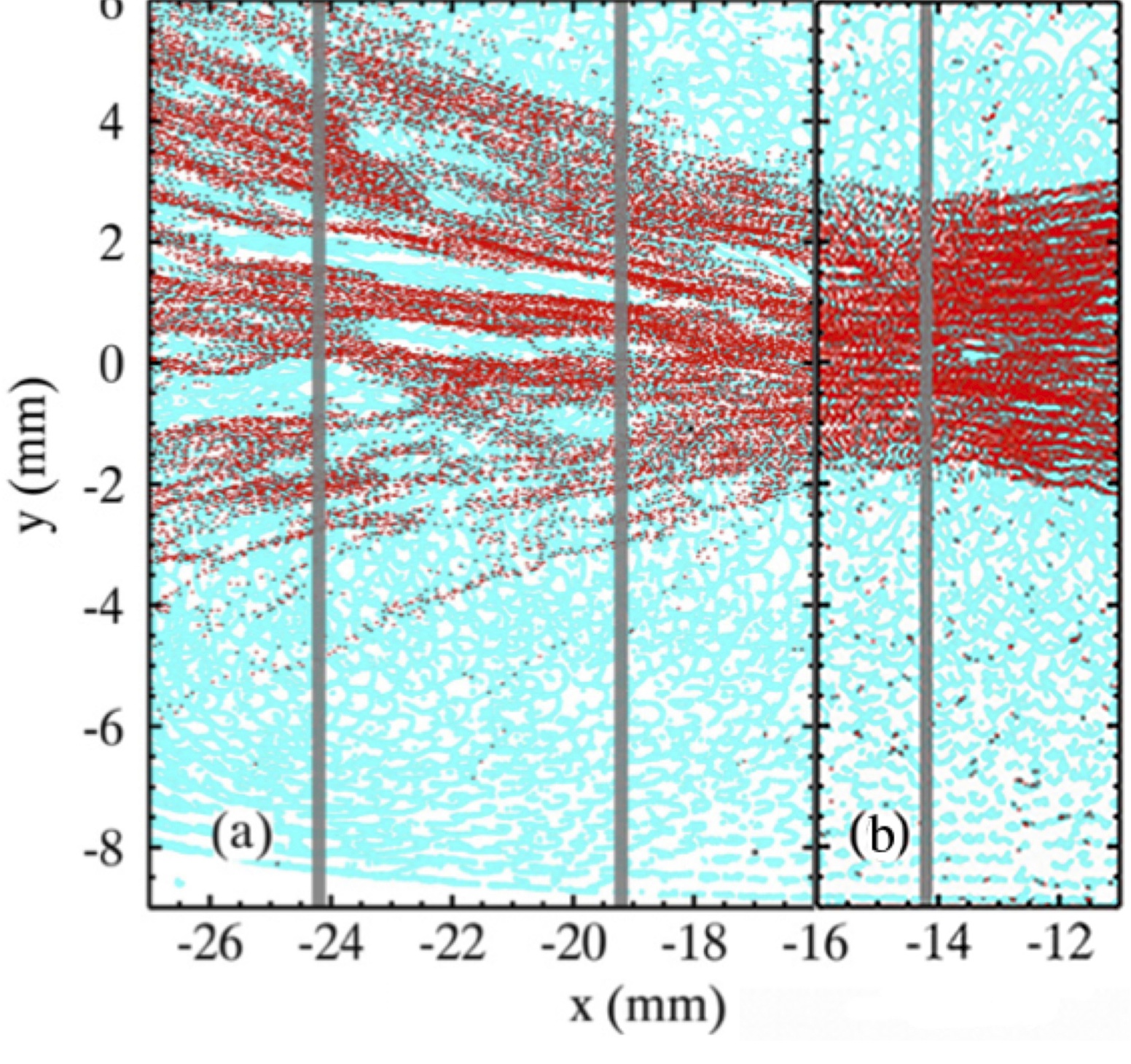}
  \caption{\label{fig:ChensLanes}(color online) Lane formation in the
    PK-3 Plus laboratory on board the International Space
    Station. Microparticles of \SI{3.4}{\micro \meter} diameter, shown
  in red/dark gray, move through a cloud of \SI{6.8}{\micro \meter} diameter
  particles. The buffer gas is argon at a pressure of
  \SI{30}{\pascal}. The crossover from lane formation (panel a) to phase
  separation (panel b) is visible. Courtesy of
  \citet{Du2012a}. \copyright IOP Publishing Ltd and Deutsche Physikalische Gesellschaft. Published under a CC BY-NC-SA licence.}
\end{figure}

\begin{figure}[t]
  \centering
  \includegraphics[width=0.5\linewidth]{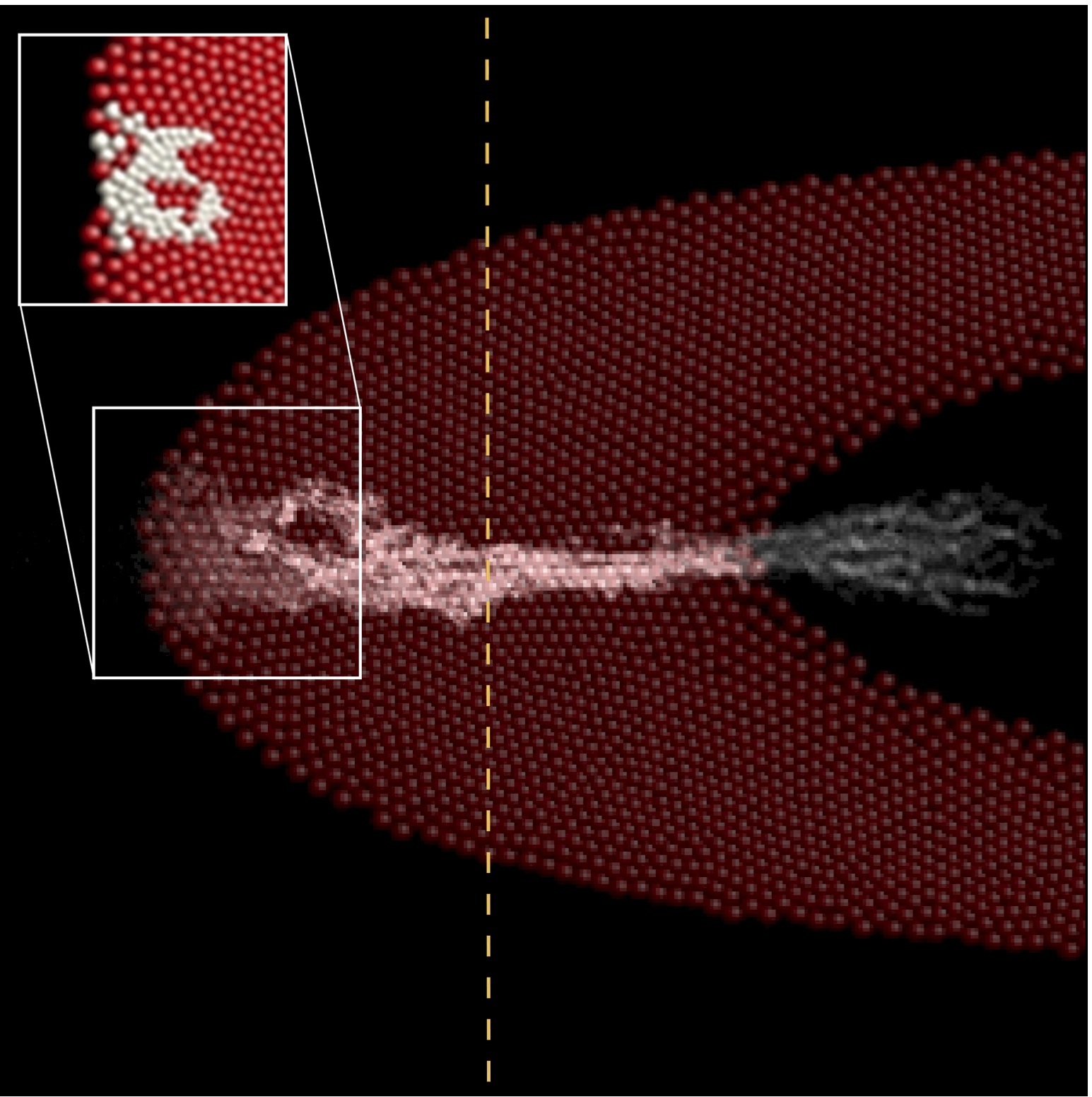}
  \caption{\label{fig:laneformation}(color online) A batch of microparticles of 3.4~$\mu$m
    diameter (white) in a cloud of particles with 6.8~$\mu$m diameter, moving towards the center of the chamber. In the
    background, the positions of the small particles are overlaid on
    the background cloud during their voyage towards the center. The
    total field of view is \SI{15 x 15}{\milli \meter}. The inset
    (field of view: \SI{3.8 x 3.8}{\milli \meter})
    shows one frame when the microparticles start penetrating into the
    background cloud, starting to form lanes. The gas pressure is
    10~Pa, input current 12~mA. In contrast to the other
    examples shown in this paper, here, the sheath width is set to
    \SI{3}{\milli \meter}, and the temperature that the microparticles
    are subjected to from the background gas is \SI{600}{\kelvin}. The
  vertical dashed line marks the approximate position where lane
  formation turns into phase separation.}
\end{figure}

Figure~\ref{fig:ChensLanes} by \citet{Du2012a} demonstrate this crossover. Small
particles of \SI{3.4}{\micro \meter} diameter are injected into a
cloud of \SI{6.8}{\micro \meter} diameter particles. First, they form
lanes, then the small particles compress into a droplet, from which
the larger particles are expelled. 

Figure~\ref{fig:laneformation} shows simulated lane formation in
argon at a pressure of 10~Pa and an input current of 12~mA. In
contrast to the other simulations in this paper, this one 
was run with a sheath width of \SI{3}{\milli \meter}, and the
temperature that the gas transfers to the microparticles was
\SI{600}{\kelvin}. The large particles, shown in red/dark gray, have a diameter of
6.8~$\mu$m, the small particles, shown in white, have a diameter of
3.4~$\mu$m, as in the experiments of \citet{Du2012a}. We use the
microparticle charges determined in \cite{Du2012a}, namely
\SI{-4500}{e} and \SI{-1900}{e}. 

As in \cite{Du2012a}, we observe all three stages --
driven movement towards the cloud, lane formation, and crossover to
phase separation --during the course of the simulation. In
Fig.~\ref{fig:laneformation}, the approximate point where the
cross-over from lane formation to phase separation occurs is marked
with a vertical dashed line. The driven movement of the small
particles towards the cloud is not visible in this figure. This stage is identical to the injection process shown in Fig.~\ref{fig:injection}. 

A difference between the simulation and the experiment is that the particles move
significantly more slowly to the center of the cloud, and the
crossover to phase separation occurs much faster. This is probably due to the lower injection speed
(see paragraph on confinement and void formation) and to the fact that
the cloud of bigger particles is partly crystallized in the
simulation. Also, there are vortices present in the simulation,
which suppress the lanes formed further away from the axis of
symmetry. Another difference is that the lane formation in the
beginning is less pronounced in the simulation, with fewer lanes of the larger
particles, probably for the same reasons as the lower speed. The
two-dimensional nature of our simulation also means that the microparticles
do not interact with any out-of-plane particles, as they
do in experiments. We will investigate this in more detail in future, three-dimensional simulations.

Next, we shall see what happens when a single fast, large particle penetrates
the cloud instead of many small ones.

\subsection{Mach cones}

\begin{figure}
  \centering
  \includegraphics[width=0.5\linewidth]{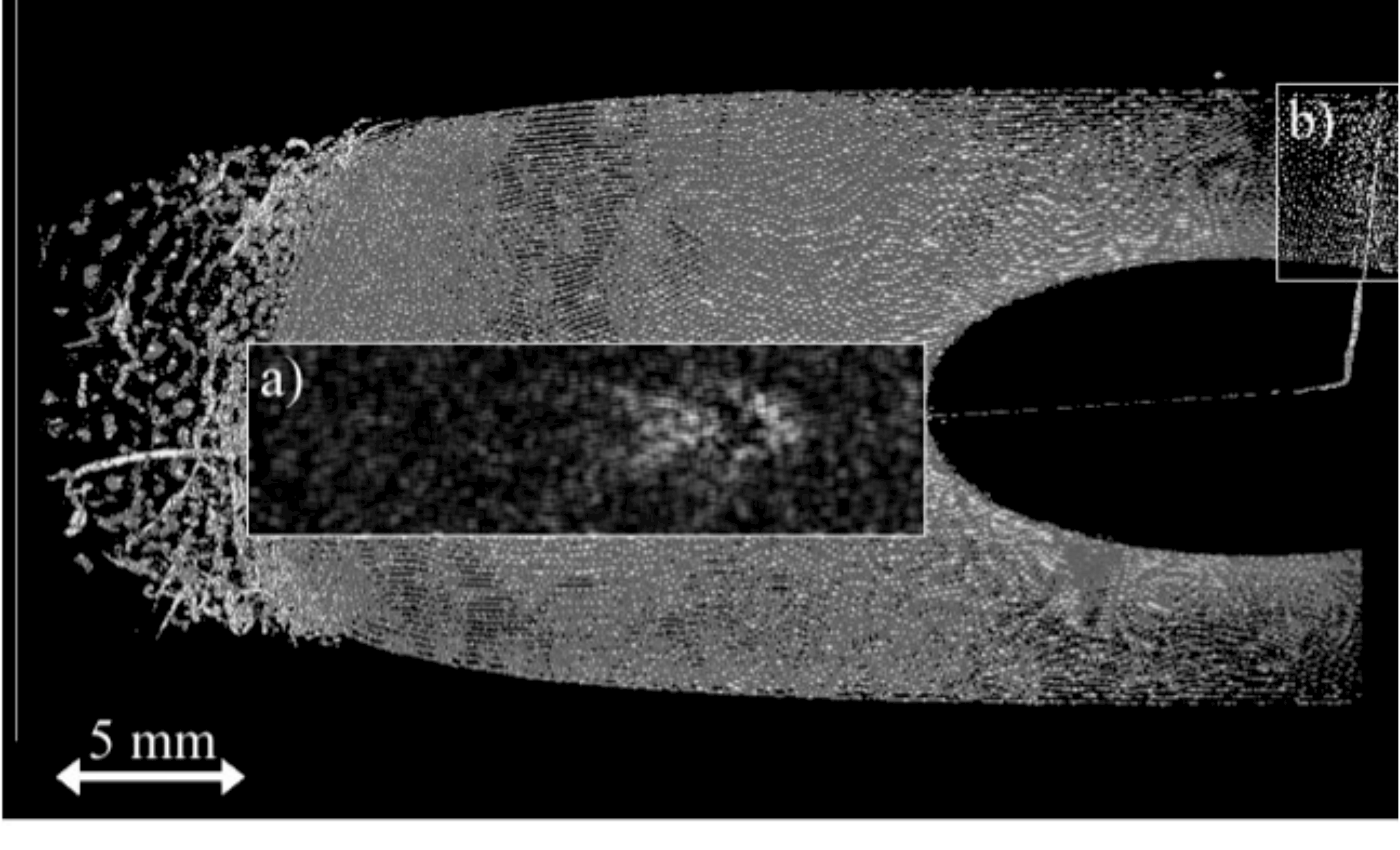}
  \caption{\label{fig:MachCone}Mach cone excited by a projectile
    moving through a cloud of \SI{2.55}{\micro \meter} diameter
    particles at a gas pressure of \SI{10}{\pascal}. The experiment
    was performed under microgravity in the PK-3 Plus laboratory on
    board the International Space Station. Inset a) shows the
    difference between two successive frames, making visible the Mach
    cone as difference in the particle densities. Inset b) is a
    superposition of 64 frames that shows the continuation of the
    trajectory upwards. Courtesy of
    \citet{Schwabe2011c}.}
\end{figure}

\begin{figure}
  \centering
  \includegraphics[width=0.5\linewidth]{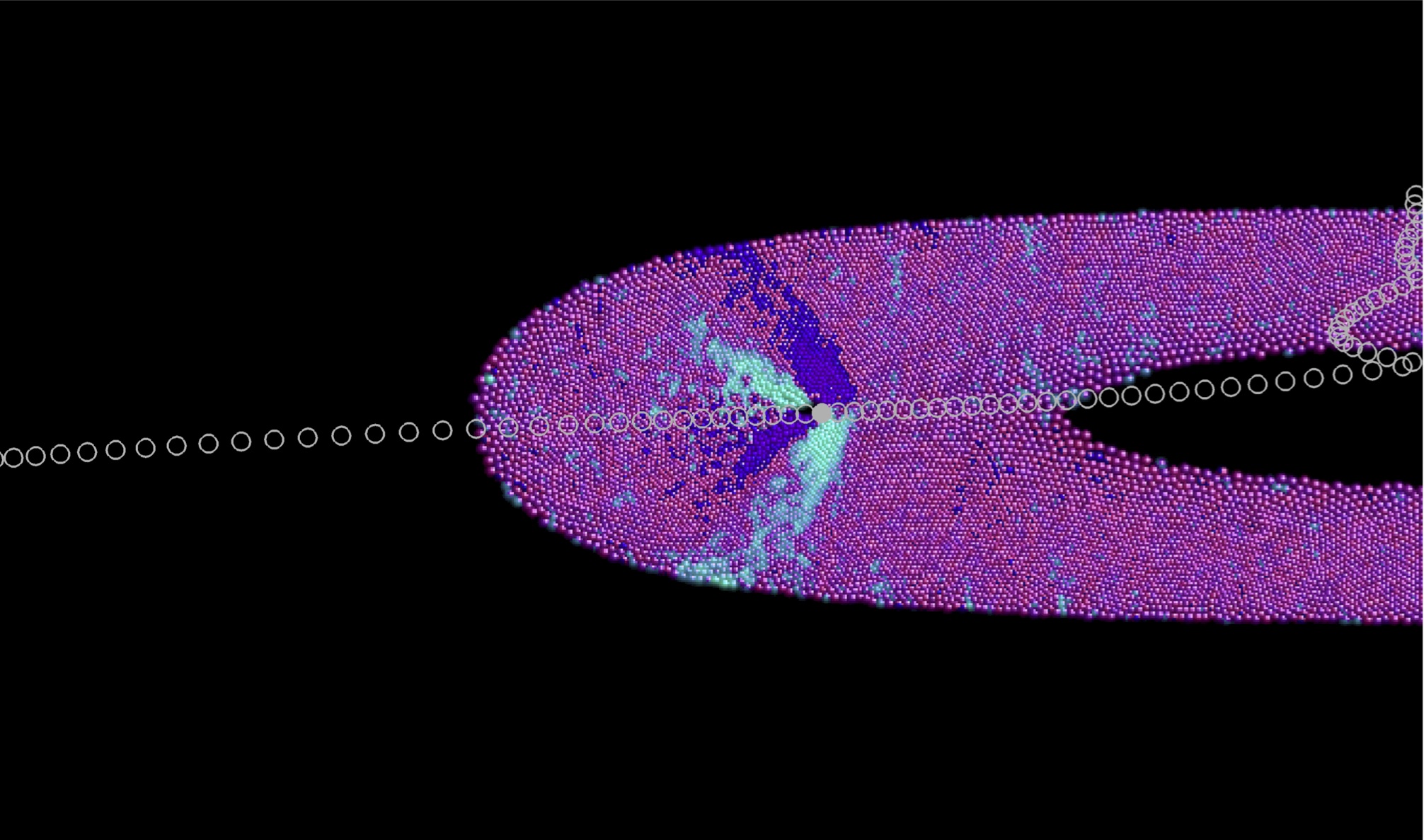}
  \caption{\label{fig:SimCone}(color online) Simulated Mach cone in a cloud of
    \SI{2.55}{\micro \meter} diameter particles at a gas pressure of \SI{10}{\pascal}. Field of view:
    \SI{33.9 x 22.0}{\milli \meter}. A projectile is dragged by an external force
    horizontally towards the chamber center and upwards. The positions of the projectile are
    shown with white circles that are spaced \SI{10}{\milli \second} apart
    in time. In the background, a single simulation frame is shown,
    with the corresponding projectile position as filled circle. The colors of the background particles indicate the vertical
    velocity of the microparticles and are scaled between
    \SI{-3}{\milli \meter \per \second} (cyan/light gray) to \SI{+3}{\milli
      \meter \per \second} (blue/dark gray). The Mach cone is clearly visible in
    the movement of the particles. The image was mirrored to the left
    to easy comparison with Fig.~\ref{fig:MachCone}.}
\end{figure}

Mach cones form when a projectile moves through a fluid faster than
the speed of sound. This is easy to achieve in complex plasmas where
the speed of sound is typically of the order of a few \si{\centi
  \meter \per \second}. In the PK-3 Plus laboratory on board the
International Space Station, sometimes bigger particles from the edge
of the plasma cloud accelerate and move through the cloud
\cite{Jiang2009,Schwabe2011c}. The reason for the sudden acceleration
is not completely understood yet -- it might be, for instance, a laser-induced rocket force \cite{Nosenko2010}. 

Figure~\ref{fig:MachCone} from \cite{Schwabe2011c} shows an example of
a Mach cone in the PK-3 Plus laboratory under microgravity conditions. The
background cloud is composed of microparticles of \SI{2.55}{\micro
  \meter} diameter. The gas pressure is \SI{10}{\pascal}. The
projectile moves from left to right, is decelerated in the void and
then moves upwards through the
microparticle cloud \cite{Zhukhovitskii2012}. 

Figure~\ref{fig:SimCone} shows a frame and the projectile trajectory from the simulation corresponding to the
experiment that we have just described. The charge of the background
particles was set to -2400 $e$, as determined in \cite{Schwabe2011c}. The projectile
    is five times larger than the background particles and has a
    charge 25 times as large. A horizontal force of
\SI{1.4e-12}{\newton} drives the projectile through the microparticle
cloud. A small vertical forces of \SI{5e-14}{\newton} is applied in
upwards direction. The white circles in Figure~\ref{fig:SimCone} mark
the position of the projectile every \SI{10}{\milli \second}. The distance
between the circles increases until the projectile reaches the
background cloud -- it accelerates in this region. Inside the cloud,
the projectile decelerates, as the decreasing distance between the
white circles shows. This is equivalent to the experimental situation,
where the projectile decelerated inside the microparticle cloud from
\SI{8}{\centi \meter \per \second} to \SI{3.7}{\centi \meter \per
    \second} \cite{Schwabe2011c}. The reason for the deceleration is
  the interaction with the background particles -- the projectile has
  to move them out of the way to be able to move. 

As in the experiment, the projectile excites a Mach cone in the background cloud. This is visible in the
colors of the microparticles (see Fig.~\ref{fig:SimCone}). The microparticles in front of the
projectile move away from
the projectile; the particles in the region above the projectile move
upwards, those in the region below it move downwards. In the region
behind the projectile, the particles move to fill the void left by the
projectile. Thus, the particles in the region above the projectile
path move downwards and the particles in the lower region move
upwards. 

 The angle of the Mach cone increases in the
 simulation as in the experiment while the projectile moves through the cloud. This is due to
 the fact that the projectile is decelerating while moving on its
 trajectory inside the cloud. 

The projectile moves upwards and finally penetrates the cloud above
the void, where it is again decelerated by the smaller microparticles. In the experiment,
  the projectile reaches the region above the chamber center and is
  consequently accelerated into the upper cloud, but its trajectory
  forms a steeper angle inside the void than in the simulation.

  The similarities between the simulation and the experiment break down when the projectile reaches the midplane
  of the chamber: In the simulation, we have applied reflecting boundary conditions at the
  vertical midplane of the chamber. In the experiment, the projectile
  moves into the right half of the chamber, in our simulation, it is
  reflected by the boundary and starts moving backwards/to the
  left. This means that it moves in opposite direction to the driving
  force, which then in turn decelerates it and turns it around
  again. In the experiment, the projectile continues a straight course
  through the upper particle cloud until it reaches the cloud edge (see
  Fig.~\ref{fig:MachCone}). 

\section{Concluding remarks}

In this paper, we have presented a new simulation of complex
plasmas. It consists of a coupled fluid simulation of the capacitively
coupled plasma chamber and a molecular dynamics simulation of the
microparticles. At the present stage, both parts of the simulation are
two-dimensional. Nevertheless, we were able to qualitatively reproduce many
phenomena of experiments with complex plasmas. For instance, the
ambipolar electric field automatically confines particles
placed at the edge of the chamber. The particles then move towards the
center, where they form a cloud with a central, particle free void. The particles
in our simulation arrange by size, with smaller particles closer
to the center than larger ones. We also observe lane formation when small
particles penetrate a cloud of larger ones, and Mach cones when very
fast projectiles penetrate the particle cloud.
 
In the future, we plan to complete the coupling between the two
simulations and include the effects of the microparticles on the
plasma. We also plan to extend the simulations to three dimensions.

\section*{Acknowledgments}
We would like to thank Emi Kawamura and Michael Lieberman
for providing the original plasma model and for offering help and advise in
applying the model. We also thank Jhih-Wei Chu, Alexei Ivlev, Sergey
Khrapak, Hubertus Thomas and Sergey Zhdanov for helpful discussions
and support. 

This research was supported by a Marie Curie International Outgoing Fellowship within the
7th European Community Framework Program. The authors acknowledge partial support from the Department of Energy, Office of Fusion Science Plasma Science Center. The PK-3 Plus project is
funded by the space agency of the Deutsches Zentrum f\"{u}r Luft- und
Raumfahrt with funds from the German Federal Ministry for Economy and
Technology according to a resolution of the Deutscher Bundestag under
grant numbers 50 WP 0203 and 50 WM 1203. 

\bibliographystyle{unsrt}

\end{document}